\documentclass{aa}

\usepackage{graphicx}
\usepackage{txfonts}
\usepackage{ulem}
\usepackage{hyperref}
\usepackage{xcolor}

\newcommand{\tbn}{$\theta_{Bn}$}

\begin{document}

   \title{A Tale of Two Shocks}


   \author{Robert F.\,Wimmer-Schweingruber \inst{1}
          \and
           Domenico Trotta \inst{2}
           \and
           Rungployphan Kieokaew \inst{3}
           \and
           Liu Yang \inst{1}
           \and
           Alexander Kollhoff \inst{1}
           \and 
           Lars Berger \inst{1}
           \and
           Patrick Kühl \inst{1}
           \and
           Stephan I. Böttcher \inst{1}
           \and
           Bernd Heber \inst{1}
           \and
           Philippe Louarn \inst{3}
           \and
           Andrey Fedorov \inst{3}
           \and
           Javier Rodriguez-Pacheco \inst{4}
           \and
           Raúl Gómez-Herrero \inst{4}
           \and
           Francisco Espinosa Lara \inst{4}
           \and
           Ignacio Cernuda \inst{4}
           \and
           Yulia Kartavykh \inst{1}
           \and
           Linghua Wang \inst{5}
           \and
           George C. Ho \inst{6}
           \and
           Robert C. Allen \inst{6}
           \and
           Glenn M. Mason \inst{7}
           \and
           Zheyi Ding  \inst{1}
           \and
           Andrea Larosa \inst{8}
           \and
           G. Sindhuja \inst{1}
           \and 
           Sandra Eldrum \inst{1}
           \and
           Sebastian Fleth \inst{1}
           \and
           David Lario \inst{9}
          }

   \institute{Institute of Experimental and Applied Physics, Kiel University,
              Leibnizstraße 11, DE-24118 Kiel\\
              \email{wimmer@physik.uni-kiel.de}
         \and
             European Space Agency (ESA), European Space Astronomy Centre (ESAC), Camino Bajo del Castillo s/n, 28692 Villanueva de la Cañada, Madrid, Spain \\
             \email{domenico.trotta@esa.int}
         \and
            Institut de Recherche en Astrophysique et Planétologie (IRAP), Toulouse, France\\
            \email{rkieokaew@irap.omp.eu}
        \and
        University of Alcalá, Alcalá de Henares, Spain\\
        \and
        School of Earth and Space Sciences, Peking University, Beijing, 100871, People's Republic of China\\
        \and
        Southwest Research Institute, San Antonio, TX, USA\\
        \and
        Applied Physics Laboratory, Johns Hopkins University, Laurel, MD, USA\\
        \and
        Istituto per la Scienza e la Tecnologia dei Plasmi, Consiglio Nazionale delle Ricerche, I-70126, Bari, Italy\\
        \and
        Heliophysics Science Division, NASA Goddard Space Flight Center, Greenbelt, MD 20771, USA
             }

   \date{Received ; accepted }

 
  \abstract
{Energetic particles in interplanetary space are normally measured at time scales that are long compared to the ion gyroperiod. Such observations by necessity average out the microphysics associated with the acceleration and transport of 10s - 100s keV particles.
}
{We investigate previously unseen non-equilibrium features that only become observable at very high time resolution, and discuss possible explanations of these features.}
{We use unprecedentedly high-time-resolution data that were acquired by the in situ instruments on Solar Orbiter in the vicinity of two interplanetary shocks observed on 2023-11-29 07:51:17 UTC and 2023-11-30 10:47:26 UTC at $\sim 0.83$ astronomical units from the Sun.}
{The solar-wind proton beam population follows the magnetic field instantaneously, on time scales which are significantly shorter than a gyro-period. Energetic particles, despite sampling large volumes of space, vary on remarkably short time scales, typically on the order of the convection time of their gyro-radius. Non-equilibrium features such as bump-on-tail distributions of energetic particles are formed by small-scale magnetic structures in the IMF.}
{High-time-resolution observations show previously unobserved microphysics in the vicinity of two traveling interplanetary shocks, including ion reflection at a current sheet, which may explain where ions are reflected in shock acceleration. }


\maketitle
%

\section{Introduction}
\label{sec:intro}


The differential flux of energetic particles in the vicinity of traveling interplanetary (IP) shocks is believed to be determined by a number of factors \citep[see e.g.,][for a review]{desai-and-giacalone-2016}. Particles are accelerated by two main mechanisms depending on the orientation of the interplanetary magnetic field (IMF) with respect to the shock normal. It has thus become customary to distinguish between (quasi-) parallel and (quasi-) perpendicular shocks. In the (quasi-) parallel case, waves upstream and downstream of the shock scatter particles back and forth across the shock and thus accelerate particles through the first-order Fermi process \cite[]{fermi-1949}. The detailed theory of this process using a diffusive approach was worked out by, e.g., \citet{jokipii-1966, fisk-1971, axford-etal-1977}. Early observations at quasi-perpendicular shocks showed that particles gain energy by (gradient) drifting along the induced electric field at the shock surface \cite[see, e.g.,]{sarris-and-vanallen1974, pesses-etal-1979, tsurutani-etal-1982}, the theory was developed by \cite{hudson-1965}. Both processes can be treated as a diffusive process and included in the particle transport equation \cite[]{parker-1965, desai-and-giacalone-2016}.  Energetic particles are thought to be accelerated out of a pool of suprathermal particles; physical and geometrical properties of the shock (such as compression ratio and angle of the shock normal) are expected to play an important role in the actual acceleration of the particles. A large body of literature exists about particles accelerated at interplanetary shocks, ranging from detailed descriptions of one single shock \cite[e.g., the November 12, 1978 quasi-parallel shock][]{kennel-etal-1984, kennel-etal-1984b, tsurutani-etal-2024} to comprehensive statistical studies of shock properties \cite[see, e.g.,][for a very recent review of shocks encountered by Solar Orbiter]{trotta-etal-2025}. \cite{kennel-etal-1985} and \cite{papadopoulos-1985} give two excellent tutorials of the physics of particle acceleration at shocks, of which \cite{kartavykh-etal-2025} provide a very recent review.

Here we report on two shocks that encountered a very dissimilar pool of suprathermal particles and were separated in time by a little under 27 hours when Solar Orbiter was at $\sim 0.83$ astronomical units from the Sun. Their properties are reported in tab.~\ref{tab:tab_event} and discussed in sec.~\ref{sec:overview}. As will be shown in sec.~\ref{sec:shock-1} and \ref{sec:shock-2} the first (quasi-parallel shock) did not appear to accelerate a significant number of particles, while the second (quasi-perpendicular shock) did have a noticeable effect. Observations around the first shock show highly correlated behavior of waves and bulk plasma at the time scale of seconds. Observations around the second shock at the same high time resolution show extremely rich fine structure and non-equilibrium distributions of energetic particles. 

We use data from Solar Orbiter, a mission of international collaboration between the European Space Agency (ESA) and NASA \cite[]{mueller-etal-2020}. Measurements of energetic particles were made by the Energetic Particle Detector's (EPD, \cite{rodriguez-pacheco-etal-2020, wimmer-schweingruber-etal-2021}) Electron-Proton Telescope (EPT) and SupraThermal Electron-Proton (STEP) sensor. EPT has four viewing directions (or telescopes). The "sun" telescope points sunward along the nominal direction of the interplanetary magnetic field (IMF), one in the opposite direction ("anti-sun"), the two others point in the "north" and "south" directions. STEP is essentially a pin-hole camera with $3 \times 5$ pixels viewing the "sunward" direction; the fields of view (FoVs) of the central $3 \times 3$ pixels add up to approximately that of the EPT "sun" telescope. The exact pointing directions of the EPT telescopes are given in \citet{rodriguez-pacheco-etal-2020}, those of the 15 STEP pixels are given in the appendix of \cite{wimmer-etal-2023}. STEP measures ions (electrons) from few keV to $\sim 80$ keV, EPT from 25 keV to 6.4 MeV (0.475 MeV). Measurements of the IMF were made by the magnetometer (MAG, \cite{horbury-etal-2020}), the solar wind was measured by the Solar Wind Analyser (SWA, \cite{owen-etal-2020}) Proton-Alpha Sensor (PAS) which measures the 3D velocity distribution function (VDF) of protons and alpha particles. 

We give an overview of the investigated time period in sec.~\ref{sec:overview} and present the behavior of the bulk plasma at the first shock in sec.~\ref{sec:shock-1}. Energetic particles observed around the second shock are presented in more detail in sec.~\ref{sec:shock-2} where we also present its substructures. We discuss the observations in sec.~\ref{sec:discussion} and wrap up in sec.~\ref{sec:concl}.

\section{Overview of the time period}
\label{sec:overview}

\begin{figure*}
    \centering
    \includegraphics[width=\textwidth]{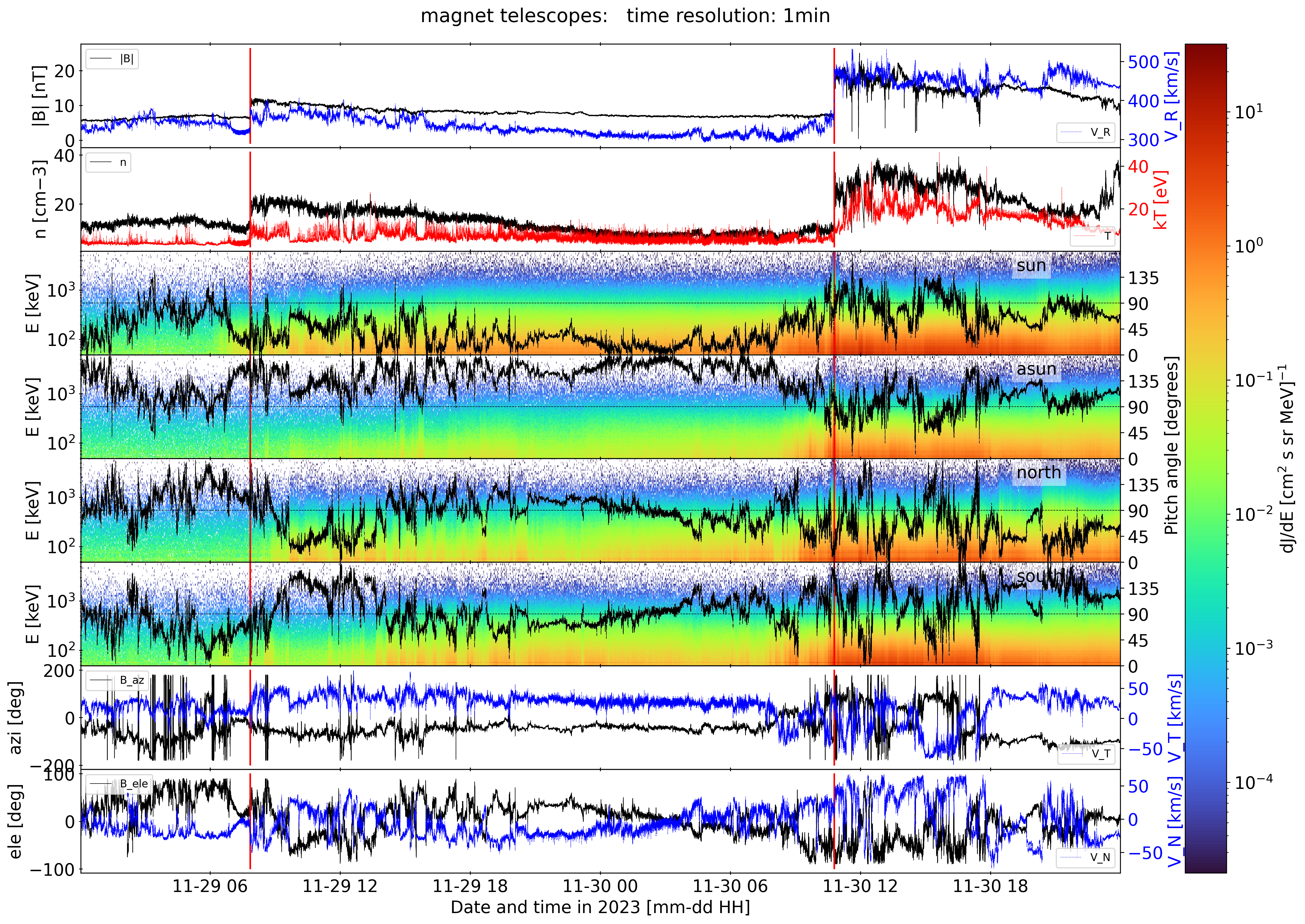}
    \caption{Overview of the time period 2023 November 29 - 30. From top to bottom: magnitude of the interplanetary magnetic field (IMF) in black and radial solar wind speed in blue ($V_R$), solar wind density (in black) and temperature (in units of kT [eV], in red), color-coded differential intensities as measured by the sun, anti-sun, north, and south telescopes of the Solar Orbiter Energetic Particle Detector (EPD) Electron-Proton Telescope (EPT). The central pitch angles of the four telescopes are overplotted with the angle information shown on the right-hand axis. The bottom two panels show the IMF azimuth and elevation angle (in black), as well as the tangential and normal components of solar wind speed ($V_T$ and $V_N$, in blue). Vertical red lines indicate the times of the two shocks.}
    \label{fig:overview}
\end{figure*}

Figure~\ref{fig:overview} presents an overview of the observations and shows, from top to bottom, the magnitude of the interplanetary magnetic field (IMF, in black) and radial solar wind speed ($V_R$, in blue) solar wind proton density (black) and temperature (red, in $kT$ energy equivalent), color-coded differential intensities as measured by the sun, anti-sun, north, and south telescopes of EPT. The central pitch angles of the four telescopes are overplotted with the angle information shown on the right-hand axis. The bottom two panels show the IMF azimuth (B\_az) and elevation (B\_ele) angle (both in black), as well as the tangential and normal components of solar wind speed ($V_T$ and $V_N$, both in blue). The two shocks are clearly visible in the IMF magnitude, they are indicated by vertical red lines at 2023-11-29 07:51:17 and 2023-11-30 10:47:26. All times are given in UTC in this manuscript. Key parameters of the two shocks are presented in tab.~\ref{tab:tab_event}. They differ slightly from the values given in the SERPENTINE data center \url{https://doi.org/10.5281/zenodo.12518015} because here we used averaging intervals which were tailored to these two specific shocks whereas the values in the SERPENTINE data base were derived using similar windows for all shock. The shock normal has been computed using the Mixed Mode 3 method~\citep[MX3][]{Paschmann2000}, and the shock speed using mass flux conservation. An ensemble of upstream/downstream averaging windows ranging from 10 seconds to 5 minutes has been used to evaluate the shock parameters~\citep[see][for a full description]{Trotta2022b}. Both shocks are supercritical as will become obvious when we present them in more detail in the following sections \ref{sec:shock-1} and \ref{sec:shock-2}. Both shocks show the "foot" and overshoots which are typical of supercritical shocks \cite[]{bavassano-catteneo-etal-1986}.

About two hours before the first shock (around 06:00), EPT's sun and south telescopes begin to see elevated particle fluxes at energies below 200 keV. Before that time, that low-energy pool population appears to be isotropic and less populated. Because the sun and south telescopes cover a similar range of pitch angles around 06:30 on 2023-11-29 they both observe the same elevated particle flux. The particle population is strongly anisotropic downstream of the first shock and streams away from the Sun. It is strongly reminiscent of a hemispheric distribution as witnessed by the oscillating intensities in the north and south telescopes that coincide with the times when the central pitch angles of these telescopes lie around 90 degrees. Particles appear to be more isotropically distributed at the shock ramp and downstream of the second shock. The second shock (2023-11-30 10:47:26) is stronger than the first and obviously has a much stronger influence on the energetic particle population which appears to be more isotropic upstream and downstream of shock \#2 than at shock \#1. 

\begin{table*}
	\centering
	\caption{Shock times and parameters inferred from Solar Orbiter observations. The parameters shown are (left to right): shock normal vector (computed using MX3), \tbn, magnetic compression ratio $\rm{r}_B$, gas compression ratio $\rm{r}$, shock speed $v_{\mathrm{sh}}$ (computed from mass flux conservation), upstream plasma beta $\beta_{\mathrm{up}}$, fast magnetosonic and Alfv\'enic Mach numbers ($\rm{M_{\rm{fms}}}$ and $\rm{M_{\rm{A}}}$, respectively).}
	\label{tab:tab_event}
	\begin{tabular}{lcccccccr} 
		\hline
		Time [UT] 2023 & $\langle \hat{\mathrm{n}}_{\mathrm{RTN}} \rangle$ &$\langle \theta_{Bn}\rangle$ [$^\circ$] & $\langle \rm{r}_B \rangle$ & $\langle \rm{r} \rangle$  & $\langle v_{\rm{sh}} \rangle \rm [km/s]$ &  $\beta_{\mathrm{up}}$& $\rm{M_{\rm{fms}}}$ & $\rm{M_{\rm{A}}}$ \\
		\hline
		 29-Nov 07:51:17 & [0.99 -0.13 -0.07] & 33 & 1.6 & 1.7 & 430 & 0.5 & 2.4 & 2.6 \\
      30-Nov 10:47:26 & [0.99 -0.03 -0.1] & 81 & 2.2  & 2 & 570 & 0.4 & 3.8 & 4.5 \\
		\hline
	\end{tabular}
\end{table*}


\section{Shock \# 1}
\label{sec:shock-1}

\begin{figure*}
    \centering
    \includegraphics[width=\textwidth]{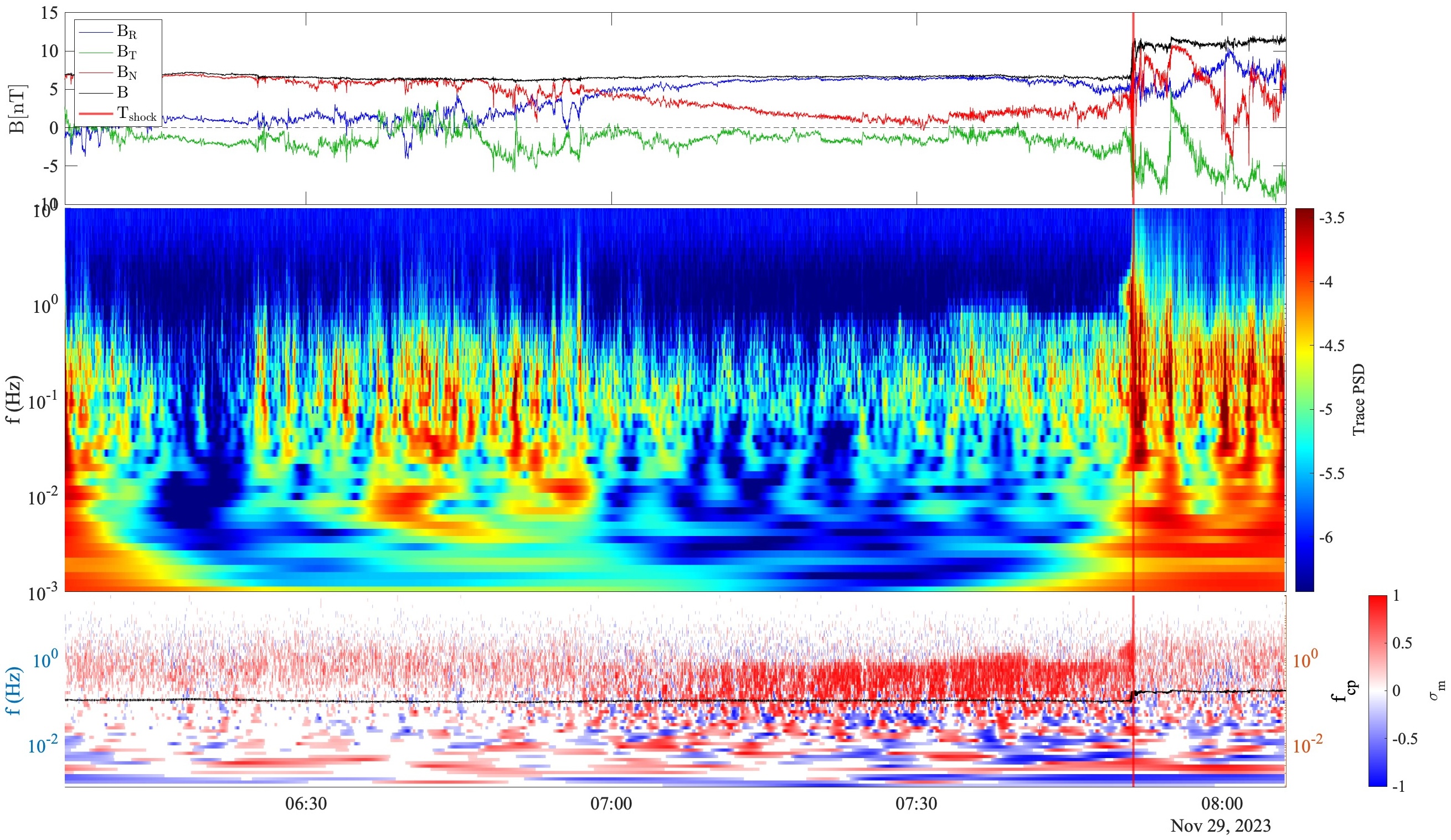}
    \caption{Analysis of the first shock shows very smooth IMF rotation just ahead of the shock (first panel), trace wavelet spectrogram of the IMF (2nd panel) showing no significant enhancement upstream of the shock transition. The normalised magnetic helicity is shown in the bottom panel (red-blue colormap), with the proton plasma frequency overplotted as a black line.}
    \label{fig:domenico-1}
\end{figure*}

Properties of the IMF surrounding shock \#1 are investigated in more detail in fig.~\ref{fig:domenico-1}. The shock is preceded by a region of low variability and a systematic rotation of the IMF between ~07:00 and the time of the shock which can easily be seen in the top panel. This indicates that Solar Orbiter was immersed in a small interplanetary coronal mass ejection (ICME) at this time, but its identification is not the topic of this paper. Nevertheless, this would explain the low level of IMF fluctuations which are also visible in the 2nd panel from the top. Analyzing the trace wavelet spectrogram in the mid-panel of fig.~\ref{fig:domenico-1}, low levels of magnetic fluctuations are observed during this time period. Closer to the shock (indicated by a vertical solid red line), starting at 7:50:00, enhanced wave activity can be seen at high frequencies ($\gtrsim 1 $ Hz). The bottom panel shows the normalized magnetic helicity, defined as $\sigma_m = \frac{2 \Im \left(\tilde{B_T^{\star}}\tilde{B_N}\right)}{|\tilde{B_R}|^2 + |\tilde{B_T}|^2 + 
|\tilde{B_N}|^2}$, where $\mathrm B$ indicates the magnetic field components, the $\tilde{}$ represents the wavelet-
transformed quantities and $\star$ represents the complex conjugation operation~\citep{Matthaeus1982}. 
Upstream of the shock, we observe a clear signature of consistently high $\sigma_m$ at ion scales, often observed in the inner heliosphere~\citep{Verniero2020ApJS}. These waves are visible as the red patches at the ion plasma frequency which is shown as the black line in the bottom panel of fig.~\ref{fig:domenico-1}. Note that these frequencies are not Doppler shifted, and a detailed investigation of such waves and their effect on the suprathermal particles upstream of shock \#1 will be part of a separate study. Here, we note that this wave activity disappears downstream of the shock, a behavior similar to what recently observed upstream of an inner heliospheric shock~\citep{Trotta2024a} and, interestingly, in a weak shock propagating in CME material as observed by \citet{trotta-etal-2024}b. In that scenario of CME-CME interaction, compatible with the one analyzed here, the low upstream level of fluctuations allows waves and shock-reflected particles to travel very far upstream of the shocks without undergoing significant scattering. This very likely explains the long-lasting ($\sim$ 1 hour) waves upstream of the first shock and the streaming particles discussed in sec.~\ref{sec:overview}.

\begin{figure}
    \centering
    \includegraphics[width=0.95\linewidth]{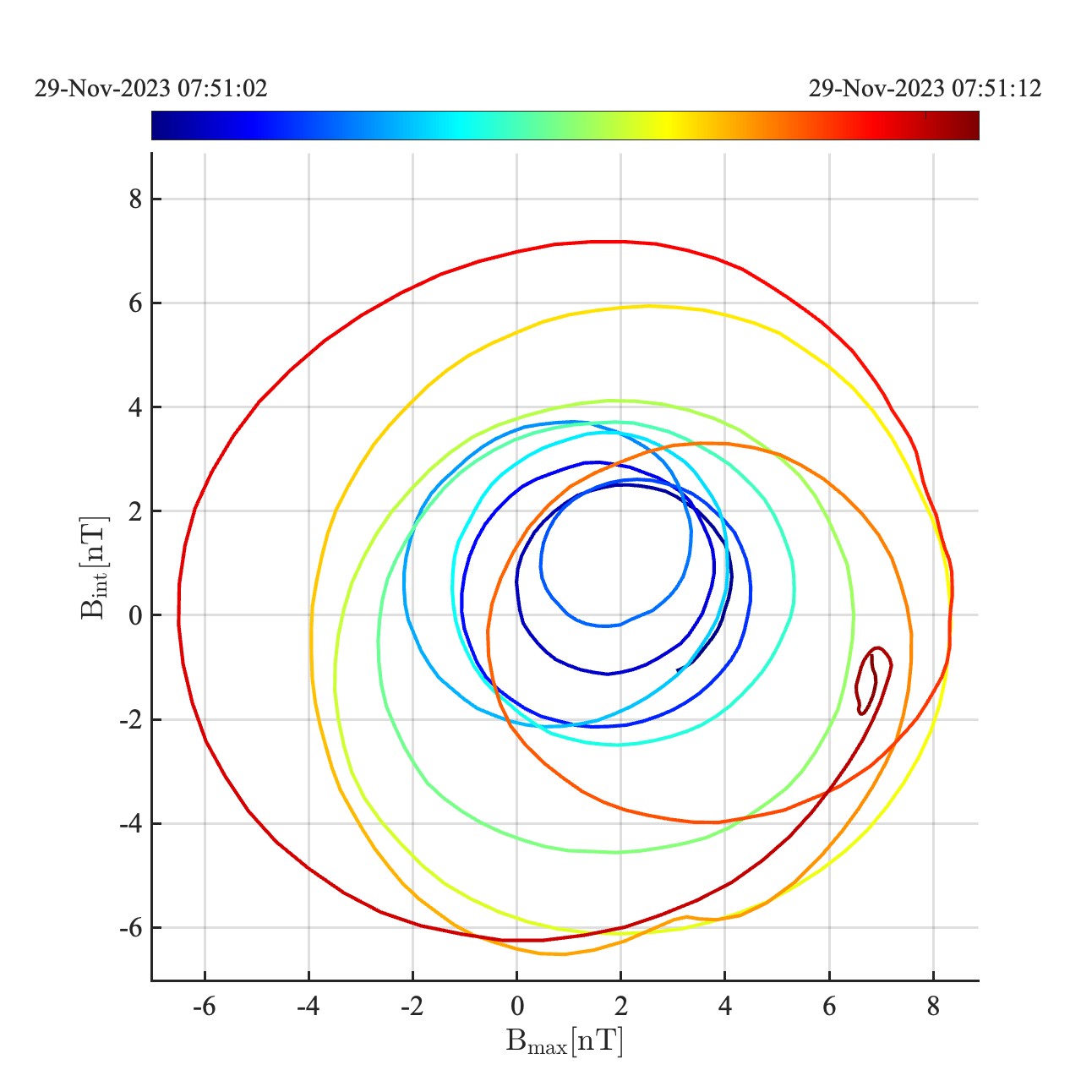}
    \caption{Hodogram of the time period 07:51:02 - 07:51:12 showing circularly polarized waves immediately upstream of the first shock.}
    \label{fig:hodogram}
\end{figure}

Further characterising the upstream fluctuations at the first shock, we performed a minimum variance~\citep[MVA;][]{Paschmann2000} analysis on the burst mode magnetic field data at 64 vec/s resolution just upstream of the first shock, between 07:51:02 and 07:51:12. In this interval, the intermediate to minimum eigenvalue ratio $\lambda_{\rm int}$/$\lambda_{\rm min}$ = 10.3, indicating that there is a well-defined minimum direction.  The relative hodogram is shown in fig.~\ref{fig:hodogram}, plotting the magnetic field in the intermediate vs.\,maximum variance directions~\citep[computed as in][]{Paschmann2000}, colored based on the measurement time. It can be seen that the waves are circularly polarised. Again, a detailed study of these interesting features will be part of a follow-up study. 

\begin{figure*}[ht]
    \centering
    \includegraphics[width=\textwidth]{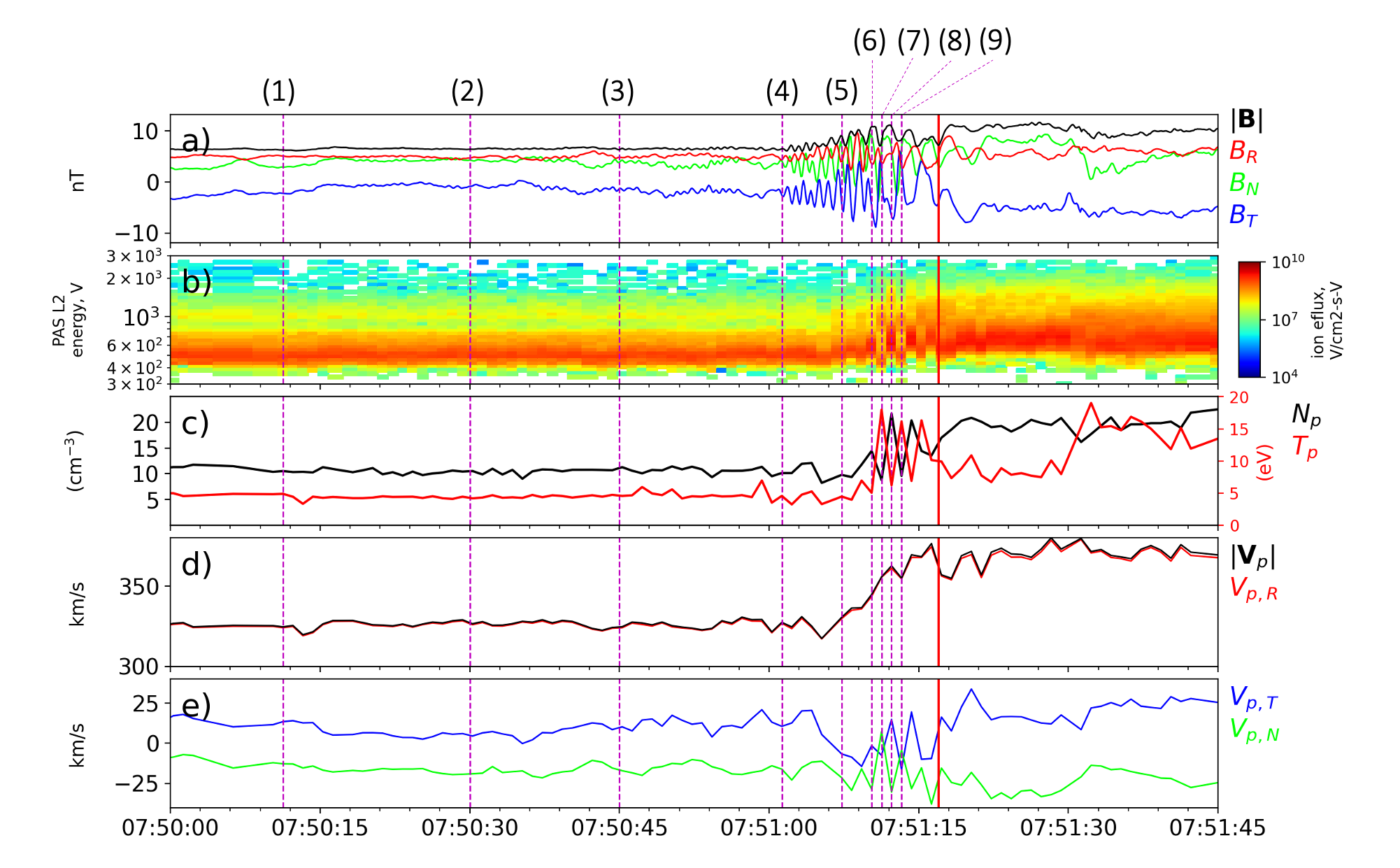}
    \caption{Solar wind properties in the vicinity of shock \#1 (indicated by the vertical red line). Panel a) shows $|B|$ in black, its RTN components in red, blue, and green. Panel b) shows the Proton-Alpha Sensor (PAS) energy-per-charge spectrum (assuming a charge of one) and shows that the proton velocity distribution is well captured in energy space. The more abundant protons are visible as the horizontal red line, alpha particles are seen as the yellow line around $10^3$ "V" before the shock. Panel c) shows proton density (temperature) in black (red). Panel d) shows radial proton velocity in red and total proton density in black. Panel e) shows the T (blue) and N (green) components of the proton velocity. Vertical dashed lines indicate the times (1) to (9) at which the reduced velocity distribution functions are shown in fig.~\ref{fig:om-2}.}
    \label{fig:om-1}
\end{figure*}

\begin{figure*}
    \centering
    \includegraphics[width=\textwidth]{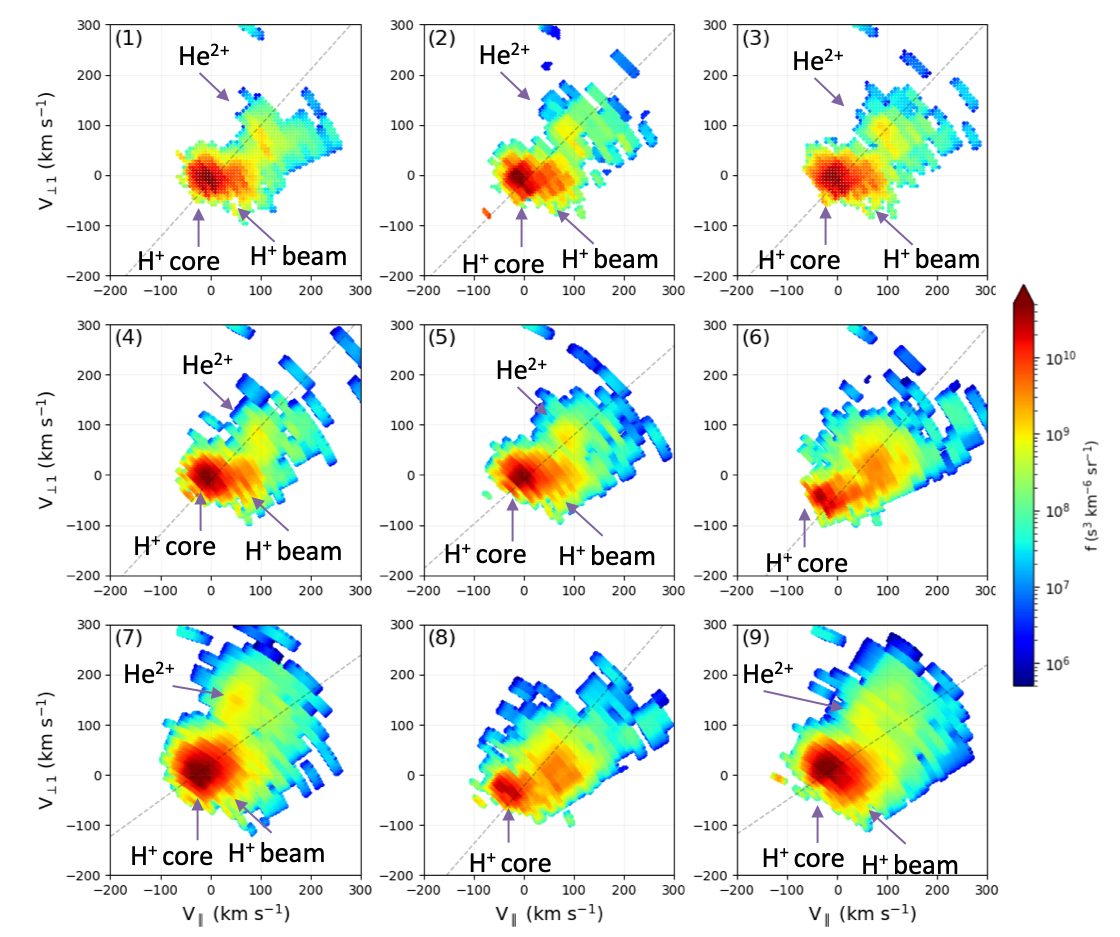}
    \caption{Reduced proton velocity distribution functions in the field-aligned coordinates (horizontal: $V_{\parallel}$; vertical:$V_{\perp 1}$ for the time periods indicated by vertical dashed lines in fig.~\ref{fig:om-1}. Proton (H$^+$) core, proton beam (along $V_{\parallel}$), and alpha (He$^{2+}$) populations are denoted. The proton bulk flow direction, where the alpha population drifts along, is marked by a grey dashed line in each plot.}
    \label{fig:om-2}
\end{figure*}

Data from the Proton-Alpha Sensor (PAS) of the SWA instrument were acquired in an experimental mode to obtain reduced velocity distribution functions (VDFs) at a cadence of one second instead of the usual four seconds. This was achieved by reducing the number of elevation angles (and energy-per-charge steps) from 9 (96) to 5 (48). To ensure that the proton peak is not lost, a full scan was performed every 100 seconds. This new operation mode can result in somewhat spiky proton moments. We compared full $E/q$-spectra at 1-s and 100-s and found that the protons were accurately tracked most of the time during the time period of interest, and that the proton VDFs were completely captured. Solar wind properties from MAG and PAS are shown in fig.~\ref{fig:om-1} with magnetic field data in panel a), an ion omnidirectional energy flux spectrogram in panel b), proton density and temperature in panel c), proton radial velocity and total velocity in panel d), and proton tangential (T) and normal (N) velocities in panel e). The spectrogram shows that PAS captured the proton VDF well, despite the restricted (but higher cadence) data product. Vertical dashed lines indicate the times for which VDFs are shown in fig.~\ref{fig:om-2}. In panel a) one can clearly see the waves emanating ahead of the shock which indicate that the shock is supercritical \cite[]{bavassano-catteneo-etal-1986}, as we mentioned in sec.~\ref{sec:overview}. Moreover, as can be seen from the VDFs collected at line 5, strong ion reflection is also found, a ``smoking gun'' of the shocks' supercriticality and rarely resolved at IP shocks because of instrumental limitations~\citep[see][]{Dimmock2023}. Interestingly, around the shock time between 07:51:00 and 07:51:15 in fig.~\ref{fig:om-1}, the proton tangential and normal velocities show fluctuations similar to those of the magnetic fields, which results in the fluctuations of the proton moments seen in panel c).

Figure~\ref{fig:om-2} shows VDFs in the field-aligned coordinates for the times delineated in fig.~\ref{fig:om-1}. The distribution function is centered at the proton bulk flow velocity for each plot. It illustrates that PAS observed strong, field-aligned proton beams already minutes ahead of shock \#1, lasting until well after passage of the shock, possibly corresponding to the persistent ion kinetic wave activity as seen in fig.~\ref{fig:domenico-1}c. For the times (1) - (3) upstream of the shock (top panels), each VDF clearly shows the proton core, and the proton beam (the sub-population of the protons that appear accelerated parallel to the magnetic field relative to the proton core). The proton beam population is rather intense, plausibly related to the ion kinetic wave activity as discussed in fig.~\ref{fig:domenico-1}c. The alpha particle (He$^{2+}$) population with its intensity about $1\%$ of that of the proton populations can be seen towards the upper right direction along the proton bulk flow direction marked by the grey dashed line. At times (4) - (6) during the shock transition (middle panels), the proton beam appears to mix with the proton core (5) and perhaps the alpha population (6). At times (7) - (9), the proton beam also appears less clear as it seems to mix with the other populations. These times correspond to the presence of the high ion-scale wave activities where the ions might be in resonance with the waves and therefore even this enhanced time resolution of PAS may not be adequate to capture the full dynamics of the proton VDF.

Strong, non-equilibrium features in the ion VDFs are visible immediately upstream of shock \#1, where we find evidence for efficient interplay between shock-reflected particles and the enhanced wave activity in the shock surroundings (see waves at 07:51 in Figure \ref{fig:om-1} and corresponding VDFs in Figure~\ref{fig:om-2}). This scenario is different from the one described by \cite{Dimmock2023}, where back-streaming particles were not found to be associated with upstream wave activity. The hodogram of the wave activity visible in Figure~\ref{fig:domenico-1} just upstream of shock \#1  at frequencies of 1 to a few Hz is shown in fig.~\ref{fig:hodogram}. The hodogram clearly shows the circular polarization of the waves which allows them to resonate with the solar wind protons. A complete characterization of the interplay between particles and waves upstream, including the long lasting cyclotron activity, will be the object of a separate study where observed non equilibrium features will be interpreted using kinetic simulations~\citep[see][]{Preisser2020}. 

\begin{figure*}
    \centering
    \includegraphics[width=\textwidth]{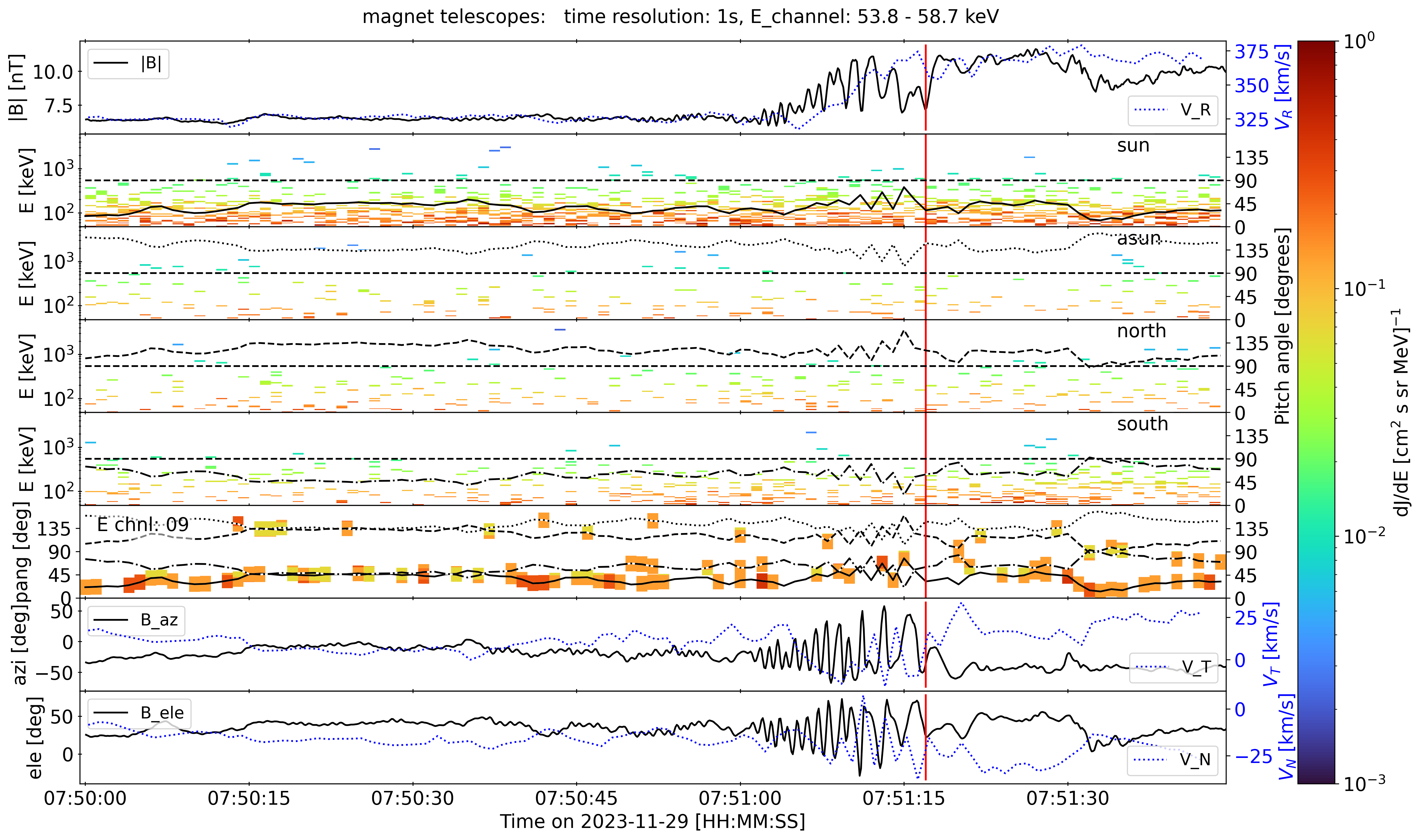}
    \caption{Close-up measurements of energetic ions at shock \#1 in a similar format as fig.~\ref{fig:overview} for the same time period as shown in fig.~\ref{fig:om-1}. Solar wind density and temperature are not shown. An additional panel is given which shows the pitch-angle distribution for $\sim 56$ keV protons (3rd panel from bottom). In it, intensity is color coded using the same color scale as for the other panels. At the prevalent IMF magnitude, the proton gyro-period is about 10 seconds, the shorter-period waves seen in the IMF are therefore not resonant with solar wind or supra-thermal protons. }
    \label{fig:shock-1}
\end{figure*}

Figure~\ref{fig:shock-1} shows the same time period as fig.~\ref{fig:om-1} with shock \#1 again indicated by a vertical red line. The format is the same as in fig.~\ref{fig:overview} with the omission of solar wind density and temperature and the addition of a panel that shows the pitch-angle distribution of $\sim$56 keV protons as observed by the four apertures of EPT. The already discussed strong wave activity is seen upstream of the shock, but the short-period ion-cyclotron waves are out of resonance with energetic and suprathermal protons. Low-energy protons are seen in the sun-telescope before and after the shock without any clear discontinuity in their flux. The third panel from the bottom shows color-coded differential intensity in the EPD energy bin $53.8 - 58.7$ keV. The flux is concentrated at pitch angles $< 90^\circ$, indicating a strongly anisotropic pitch-angle distribution. Thus, despite the very interesting kinetic physics (affecting the thermal (bulk and beam) solar wind) visible at shock \#1, it apparently did not strongly affect energetic protons (i.e., with energies greater than $\sim 25$ keV). Nevertheless, the flux of energetic protons is highly anisotropic. The flux of electrons in this event is too low to be measurable by EPT at these time scales and is therefore not shown. Based on the properties of this shock (see tab.~\ref{tab:tab_event}), we would expect it to accelerate both electrons and ions. \cite{tsurutani-and-lin-1985} found that  a minimum shock speed along the upstream IMF of 250 km/s is needed for a shock to accelerate $\ge 2$ keV electrons and $\ge 47$ keV ions. Shock \#1 clearly exceeds this limit.

As discussed above, shock \#1 is preceded by a small CME and its low level of fluctuations in the IMF. This likely also contributes to the low acceleration efficiency of this shock because the low level of upstream turbulence (inside the CME) may not be sufficient to scatter particles back towards the shock. 

\section{Shock \# 2}
\label{sec:shock-2}

\subsection{Overview of the second time period}

\begin{figure*}
    \centering
    \includegraphics[width=\textwidth]{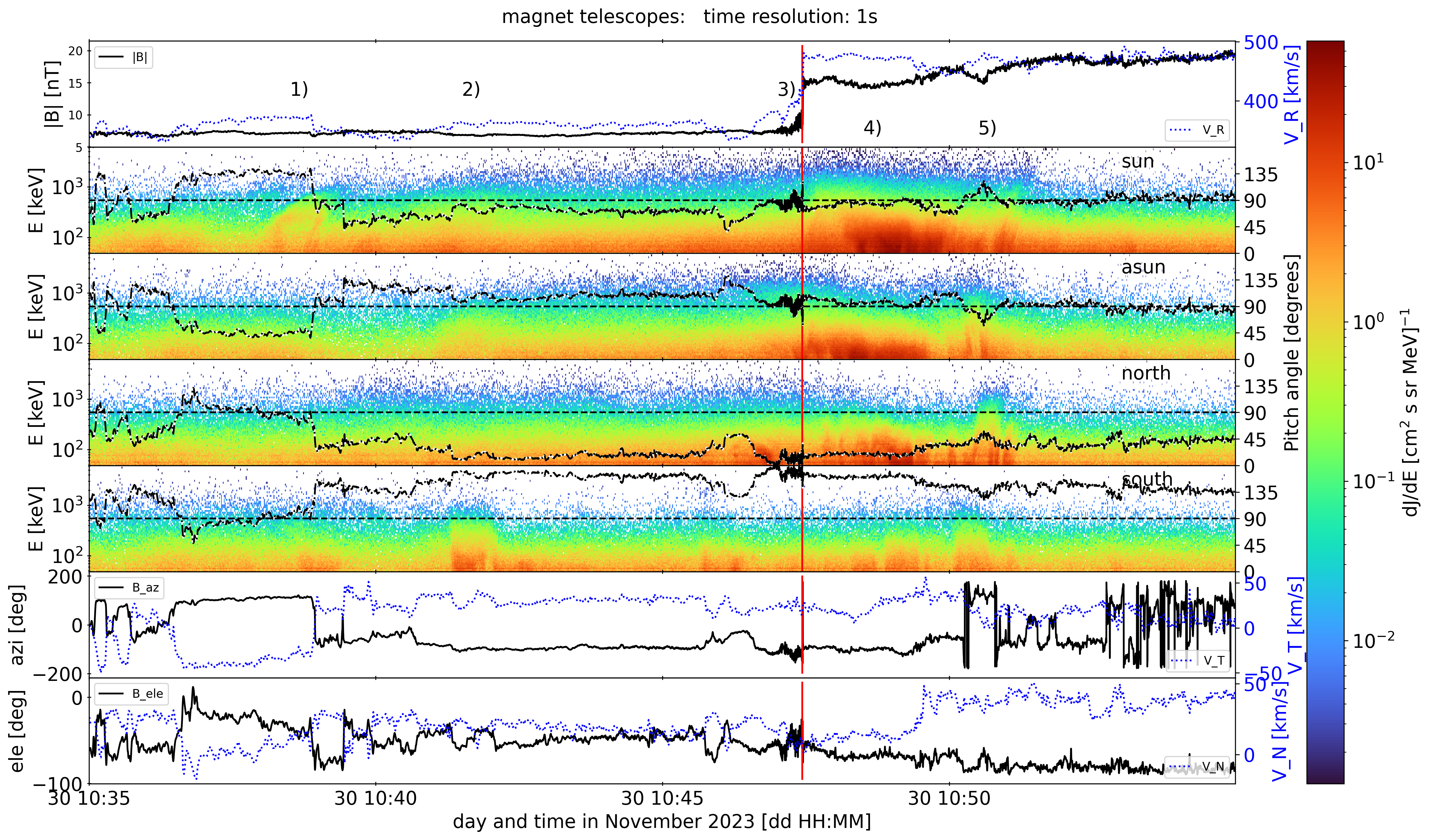}
    \caption{Overview of the second time period surrounding second shock in the same format as fig.~\ref{fig:overview}. The five interesting time periods which are described in the text are indicated by 1) -- 5) in the topmost panel.}
    \label{fig:P2}
\end{figure*}

The time period surrounding shock \#2 is shown in fig.~\ref{fig:P2} and follows the same format as fig.~\ref{fig:overview}. Five especially interesting time periods are indicated by 1) - 5) in the topmost panel and will be discussed in subsections \ref{subsec:TP1} - \ref{subsec:TP5}. Figure ~\ref{fig:overview} shows that the fluxes upstream of shock \#2 (again indicated by the vertical red line) are much more isotropic and considerably higher than at shock \#1. The time period between shock \#1 and about 2023-11-29 20:00 (see fig.~\ref{fig:overview}) is structured by many magnetic discontinuities with sudden changes in particle intensity in the different telescopes. Starting around that time IMF fluctuations are reduced and a smooth rotation sets in which could indicate that Solar Orbiter was located in a second ICME. This changes a few hours upstream of the second shock (around 2023-11-30 08:00) when fluctuations increase and the particle intensities appear more anisotropic again. Just how anisotropic and dynamic they are can be seen in fig.~\ref{fig:P2}. A highly non-equilibrium feature can be seen in the sun-ward-pointing telescope in time period 1), a sudden intensity increase is seen in the southward telescope without any apparent accompanying increase in the other telescopes at 2). Strong velocity dispersion features can be seen ahead of the shock in the anti-sun and north telescopes at 3). After the shock, significantly more disturbed conditions were observed around 4) and 5) with obvious velocity-dispersion features and their "inverses" clearly visible in the sun, anti-sun, and north telescopes. We discuss the five time periods in more detail in the following subsections.

\subsection{Time period 1)}
\label{subsec:TP1}

\begin{figure*}
    \centering
    \includegraphics[width=\textwidth]{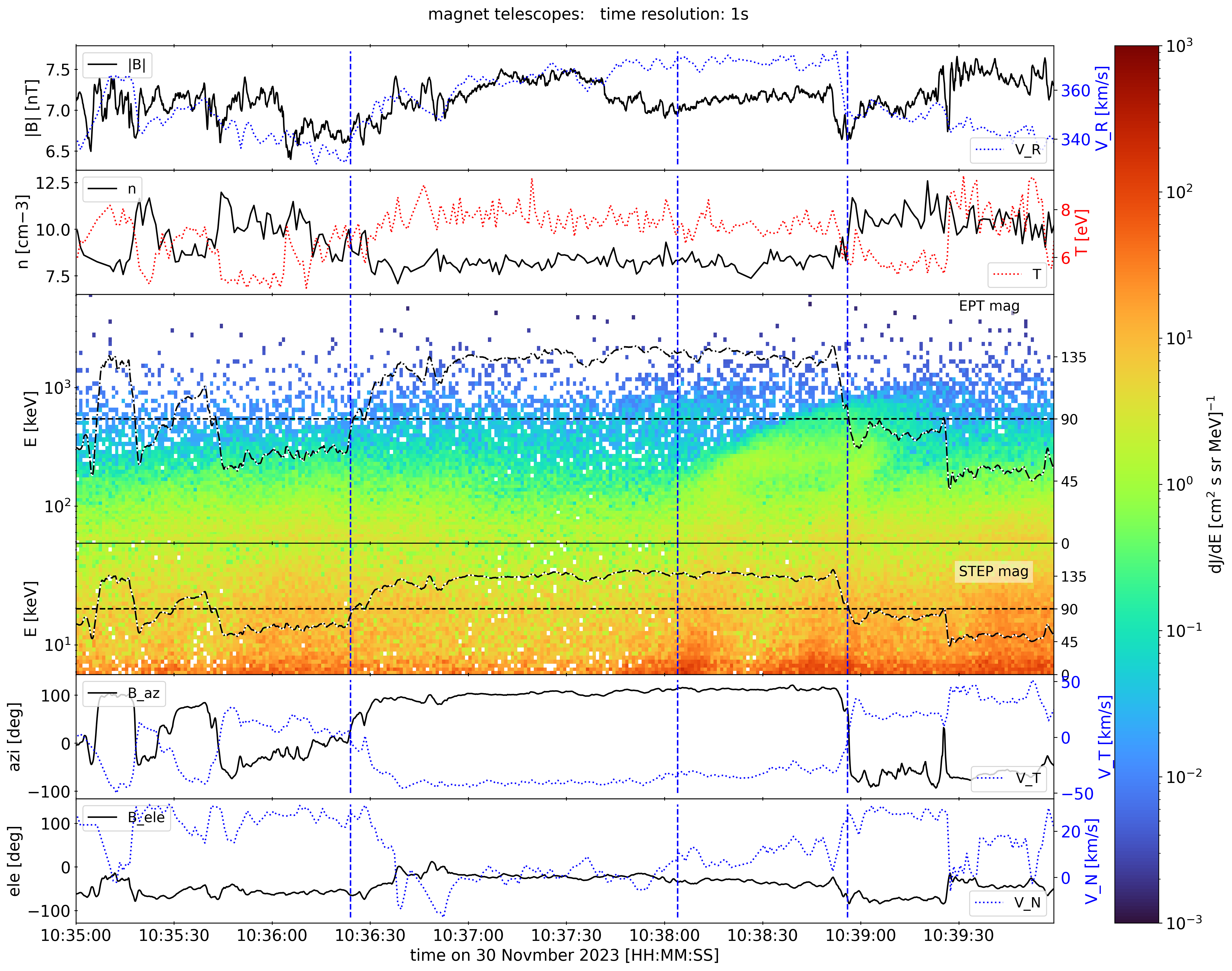}
    \caption{Time period 1) with its unstable distribution function can be seen between 10:38:00 and 10:39:00. The figure follows the usual structure but does not show all directions. It shows the sunward EPT ans STEP directions. Vertical blue dashed lines indicate the three magnetic discontinuities discussed in the text. The "loop" is visible between 10:38 and just beyond 10:39 between the two blue lines. Measurements of this time period with the STEP sensor are shown in fig.~\ref{fig:STEP_TP1} and clearly indicate that this time period was very anisotropic even within the narrow STEP FoV.}
    \label{fig:P2.1}
\end{figure*}

Figure~\ref{fig:P2.1} zooms in to time period 1 in a similar format as fig.~\ref{fig:overview} but shows only the sunward telescope of EPT (3rd panel from the top) and, in addition, data from the central $3 \times 3$ pixels of STEP (4th panel from the top). A "loop"-like feature is clearly seen between 10:38 and just beyond 10:39 between the second and third vertical blue lines (counted from the left). It clearly extends from STEP energies to EPT energies but is in fact extremely anisotropic, as can be seen in fig.~\ref{fig:STEP_TP1} in appendix \ref{app:step}. The velocity distribution between the second and third vertical dashed blue lines is obviously unstable, but lasts only for about one minute. It ends at the magnetic discontinuity (3rd blue line), but extends beyond that at higher energies. Nevertheless, it persists for several proton gyro-periods ($\tau_p \sim 10$ seconds), thus providing a limit on the growth rate of instabilities which drive the system towards equilibrium. It is worth noting that such features would not have been recognized with traditional particle instruments because of their much lower time and energy resolution. 

\subsection{Time period 2)}
\label{subsec:TP2}

\begin{figure*}
    \centering
    \includegraphics[width=\textwidth]{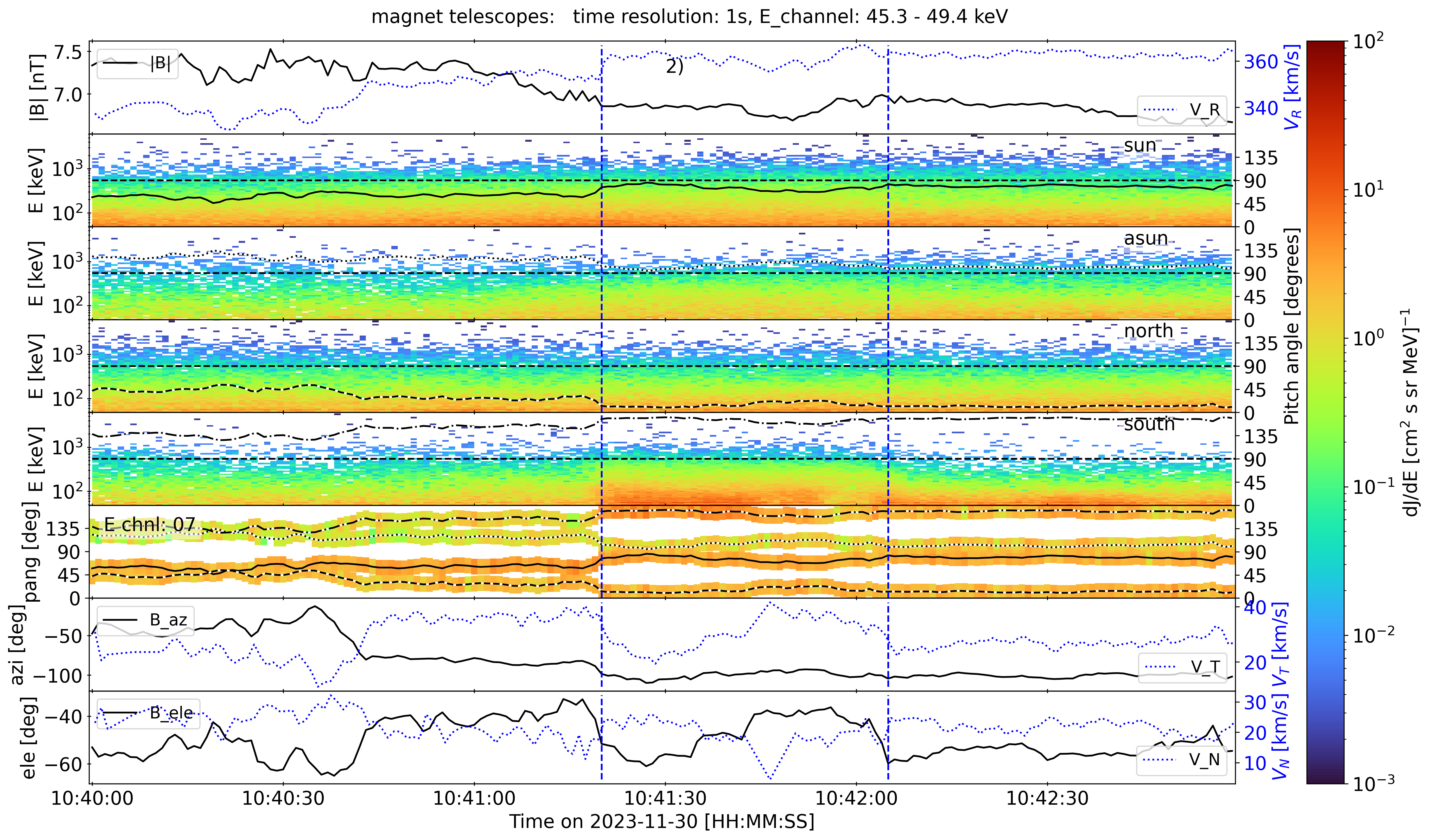}
    \caption{Time period 2 with anisotropic distribution function can be seen between 10:41:20 and 10:42:05, as indicated by the vertical blue dashed lines. The format is the same as in fig.~\ref{fig:shock-1}.}
    \label{fig:P2.2}
\end{figure*}

Time period 2) is delineated by two vertical dashed blue lines at 10:41:20 and 10:42:05 in fig.~\ref{fig:P2.2} which is in the same format as fig.~\ref{fig:shock-1}. Before 10:41:20 the energetic particle flux appears hemispheric with a higher flux for pitch-angles below 90$^\circ$ than above that. This changes abruptly at 10:41:20 when an anti-parallel beam appears at large ($\sim 180^\circ$) pitch angles and at low energies ($E < 250$ keV). Remarkably, this is not visible at pitch angles only slightly larger than $90^\circ$. This "beam" or flux enhancement in the south telescope persists for about 45 seconds, i.e., for several gyro-periods after which the pitch-angle distribution appears more isotropic at $\sim 50$ keV, the approximate energy that corresponds to energy channel 7 for which the pitch-angle distribution is shown in the sixth panel from the top. Inspection of panels 2 - 5 shows that the flux in the other telescopes appears to be unaffected during this time period, only the southward-pointing telescope measures this enhanced flux. The central pitch angles of the four telescopes change by $\sim 10$ degrees at the beginning of time period 2), but do not change significantly at its end. This indicates that the observed "beam" must be narrow and short-lived - it fades out and there is no appreciable difference in fluxes in the four telescopes at the end of time period 2). This hints at highly dynamic acceleration at a remote location, similar to what has been described by \cite{trotta-etal-2023} and \cite{yang-etal-2023, yang-etal-2024} using data from Solar Orbiter and \cite{lario-etal-2008} using data from the Advanced Composition Explorer (ACE).

\subsection{Time period 3)}
\label{subsec:TP3}

\begin{figure*}
    \centering
    \includegraphics[width=\textwidth]{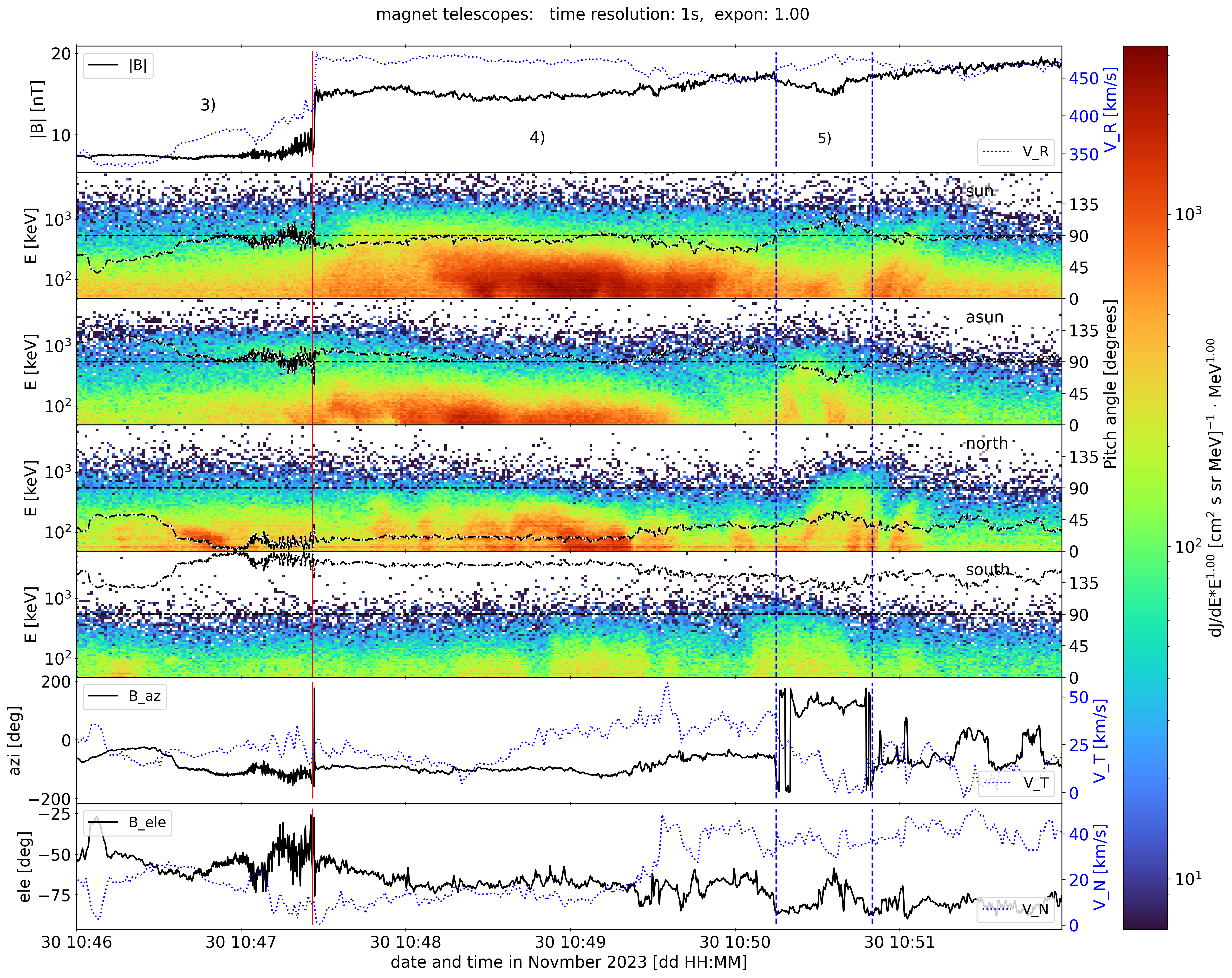}
    \caption{EPT measurements of suprathermal protons between 2023-11-30 10:46 and 10:52 show significant substructure at time scales comparable to and shorter than the proton gyroperiod of 4 - 5 seconds. The format is similar to that of fig.~\ref{fig:P2}, but intensities have been multiplied by energy to more clearly show the substructures. Time periods 4) and 5) are also indicated. Shock \#2 is indicated by the vertical red line, the beginning and end of the likely magnetic switchback that causes the dynamical behavior in time period 5) are indicated by vertical dashed blue lines.}
    \label{fig:detail-shock-2}
\end{figure*}

Figure~\ref{fig:detail-shock-2} is in the usual format, but differential intensities have been multiplied by energy to more clearly show the many substructures seen in time periods 3) - 5). The shock is indicated by the vertical red line, the vertical dashed blue lines are discussed in subsection~\ref{subsec:TP5}. The overshoot features typical of supercritical shocks \cite[]{bavassano-catteneo-etal-1986} are also present but not as obvious as for shock \#1 because of the more compressed $x$ axis, the larger $\theta_{Bn}$ and the higher shock speed in the spacecraft rest frame (see Table~\ref{tab:tab_event}). The feature at time period 3) can be seen in the northward-pointing telescope between 10:46:40 and 10:46:57 and appears to show velocity dispersion. It begins at about the time where fluctuations in the IMF begin to be visible and probably only appears to end because the energy of this "beam" was too low to be measured by EPT. As STEP does not have a telescope in the north direction, we could not measure this putative extension. This dispersive feature is seen at small pitch angles, i.e., points along the IMF and precedes the shock by approximately one minute. We interpret this as a consequence of irregular injection at a remote location along the shock surface, similar to what has been reported by \cite{trotta-etal-2023} and \cite{yang-etal-2023, yang-etal-2024}, but more pronounced. Additional dispersive features are seen at similarly low energies in the anti-sun telescope immediately ahead of the shock during the prominent IMF fluctuations (even when the pitch angle scanned by this telescope is around 90 deg).

\subsection{Time period 4)}
\label{subsec:TP4}

\begin{figure*}
    \centering
    \includegraphics[width=\textwidth]{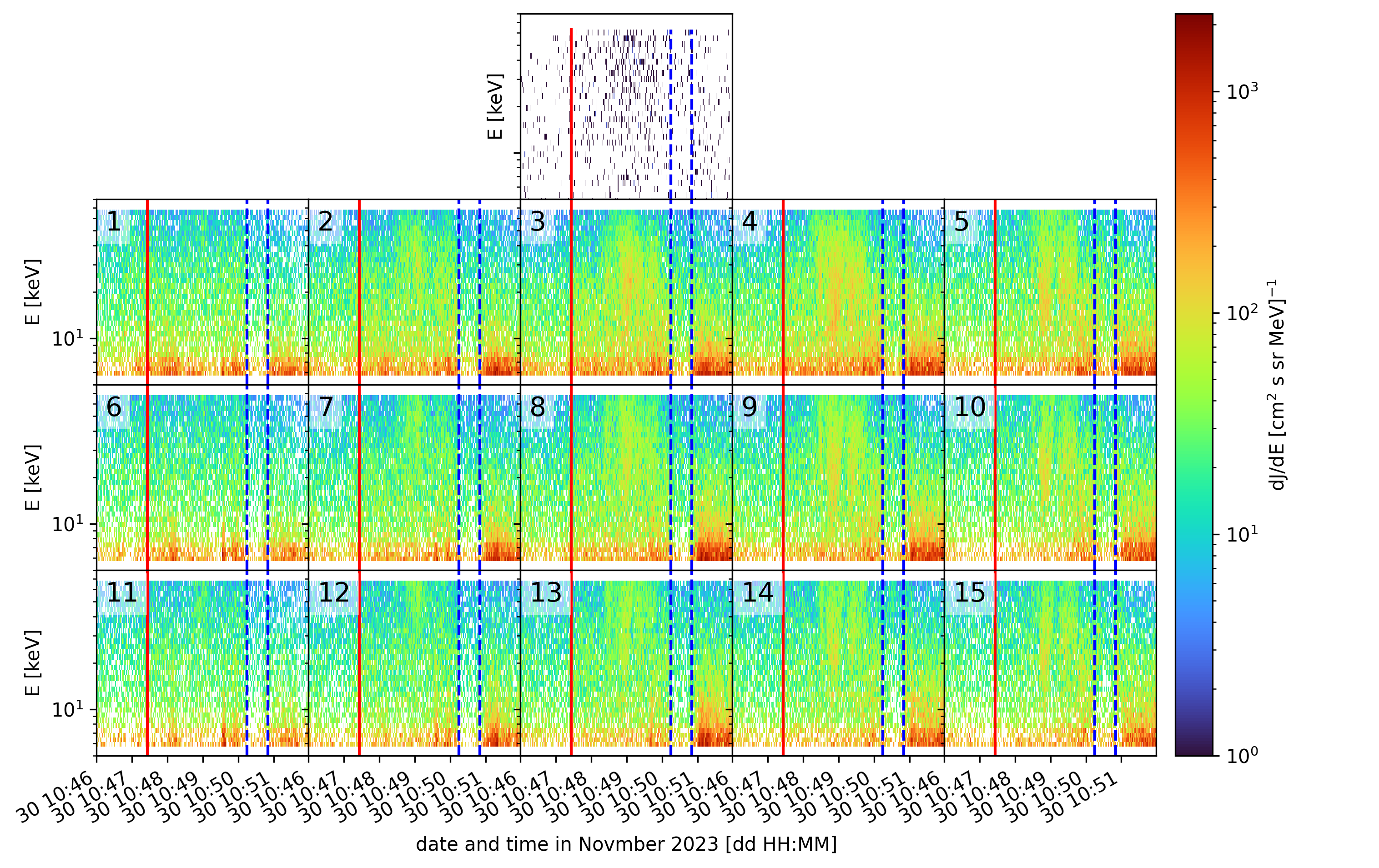}
    \caption{STEP measurements during time periods 4) and 5). Pixel numbers are given in the upper left corners of each panel, the central top pixel is the background pixel. The red line indicates shock \#2, the two dashed blue lines indicate time period 5. See text for discussion.}
    \label{fig:STEP_TP4_TP5}
\end{figure*}

Time period 4 is located downstream of the shock and the particle flux is clearly anisotropic. Despite very similar pitch angle coverage around 90$^\circ$ the sun and anti-sun telescopes see different intensities and time variations. If one looked at this time period with a more conventional time resolution of, say, one minute, and with considerably less energy resolution, one would interpret these measurements as due to a hemispherical distribution with a filled sunward and empty anti-sun hemisphere (when defining sunward as pointing towards the Sun along the magnetic field). At the higher time and energy resolution offered by EPD, this view can not be upheld, especially not when considering the measurements in the 15 individual STEP pixels which are shown in fig.~\ref{fig:STEP_TP4_TP5}.

Figure~\ref{fig:STEP_TP4_TP5} shows measurements with EPD's STEP sensor's individual pixels for the same time period as shown in fig.~\ref{fig:detail-shock-2}. Pixel numbers are indicated in the upper left corners of each panel, the panel in the top center is the background pixel and shows only individual counts. All other pixels measured considerably higher fluxes during these time periods. The vertical red lines indicate the time of shock \#2, the dashed blue lines indicate time period 5) which is discussed in sec.~\ref{subsec:TP5} below. Comparing intensities between the red and the first blue line in the different pixels shows just how anisotropic time period 4 was and that a simple-minded interpretation of a filled hemisphere does not correspond to the much more complex geometry. While pixel 4 sees the highest flux, pixels 1, 6, and 11 see virtually no elevated flux. At the time when the flux at lowest energies is most intense in EPT's sun telescope (see second panel in fig.~\ref{fig:detail-shock-2} around 10:48:30), not all of STEP's pixels see such an intensity maximum. The maximum is seen in pixel 4, intensities drop off towards surrounding pixels. The bore-sight of pixel 8 coincides with that of EPT's sunward-pointing telescope, the surrounding pixels more or less cover EPT's sunward field of view. It is obvious that the directional flux was not isotropic within EPT’s field of view. EPT’s conical fields
of view span $\pm 15^{\circ}$, the 15 pixels of STEP span  $28^{\circ} \times 54^{\circ}$ \citep{rodriguez-pacheco-etal-2020}, their pointing directions are given in \cite{wimmer-etal-2023}.

\subsection{Time period 5)}
\label{subsec:TP5}

Time period 5) is indicated by vertical dashed blue lines in Figs.~\ref{fig:detail-shock-2} and \ref{fig:STEP_TP4_TP5} and exhibits unusually dynamic behavior. Many dispersive features can be seen in all telescopes except the southward-pointing one. Fluxes are low in the latter which may explain why no dispersive features are seen by it. As fluctuations in the IMF and therefore also the pitch angle of the individual telescopes are small, the only obvious large-scale feature is a sudden jump in direction seen at 10:50:10 and back again around 10:50:50. The structure is reminiscent of a switchback \cite[]{dudok-dewit-etal-2020, zank-etal-2020} and indicated by the dashed blue lines. The dispersive features are not limited to this time period but begin about ten seconds ahead of it and last until just after 10:51 on November 30, 2023. In the anti-sun telescope the features are as expected for normal velocity dispersion, particles with higher energies arrive before those of lower energies. This is not the case for the sun and north telescopes in which higher-energy particles are measured after the lower-energy particles of the same feature. 

\section{Discussion}
\label{sec:discussion}

While the study of collisionless shocks in the heliosphere is a mature field \cite[e.g.,]{kennel-etal-1985, papadopoulos-1985}, new measurements at unprecedented time and energy resolution in the inner heliosphere are providing new and exciting insights into the microphysics of the processes associated with particle acceleration, so far mostly tested at the Earth's bow shock only.
The two events analyzed here provide an excellent opportunity to study a variety of processes of fundamental importance for energy conversion in space and astrophysical plasmas. In particular, shock \#1 shows a plethora of small-scale features, namely shock--induced upstream fluctuations and strongly out-of-equilibrium velocity distribution functions that underline the importance of ion kinetics in reconstructing how energy is converted at collisionless shocks. Such processes, widely addressed at Earth's bow shock~\citep[e.g.,][]{Schwartz2022}, remain elusive at IP shocks due to technical challenges, for which Solar Orbiter is advancing the state-of-the-art~\citep[see][for example]{Dimmock2023}. Further, as efficient proton acceleration is observed at shock \#2, this study elucidates how preceding events could play a role in pre-conditioning plasma for further acceleration. Such multiple events are more common and may be important in conditions of solar activity maximum~\citep[see][for an overview]{Trotta2025b}. However, to pin-point the role of preceding events in the generation of particle distributions that can be easily re-accelerated to higher energies, input from numerical modeling would be crucial~\citep[see][]{Nyberg2024}.  
The contrasting observation that shock \#1 had virtually no effect on energetic particles underlines this point because it was preceded by a small CME with its low level of fluctuations. This compounds the problem for particles to cross the 90$^\circ$ pitch angle to be reflected back into the shock for further acceleration. As was pointed out by \cite{tsurutani-etal-2024}, it is well known that there is no resonance with particles at 90$^\circ$ in the energy range considered in this study \cite[]{horne-etal-2003}. The low level of fluctuations inside the CME further enhances this problem. Thus, this shock (\#1) is an extreme, limiting example of a shock that does not accelerate particles.

\begin{figure*}
    \centering
    \includegraphics[width=\linewidth]{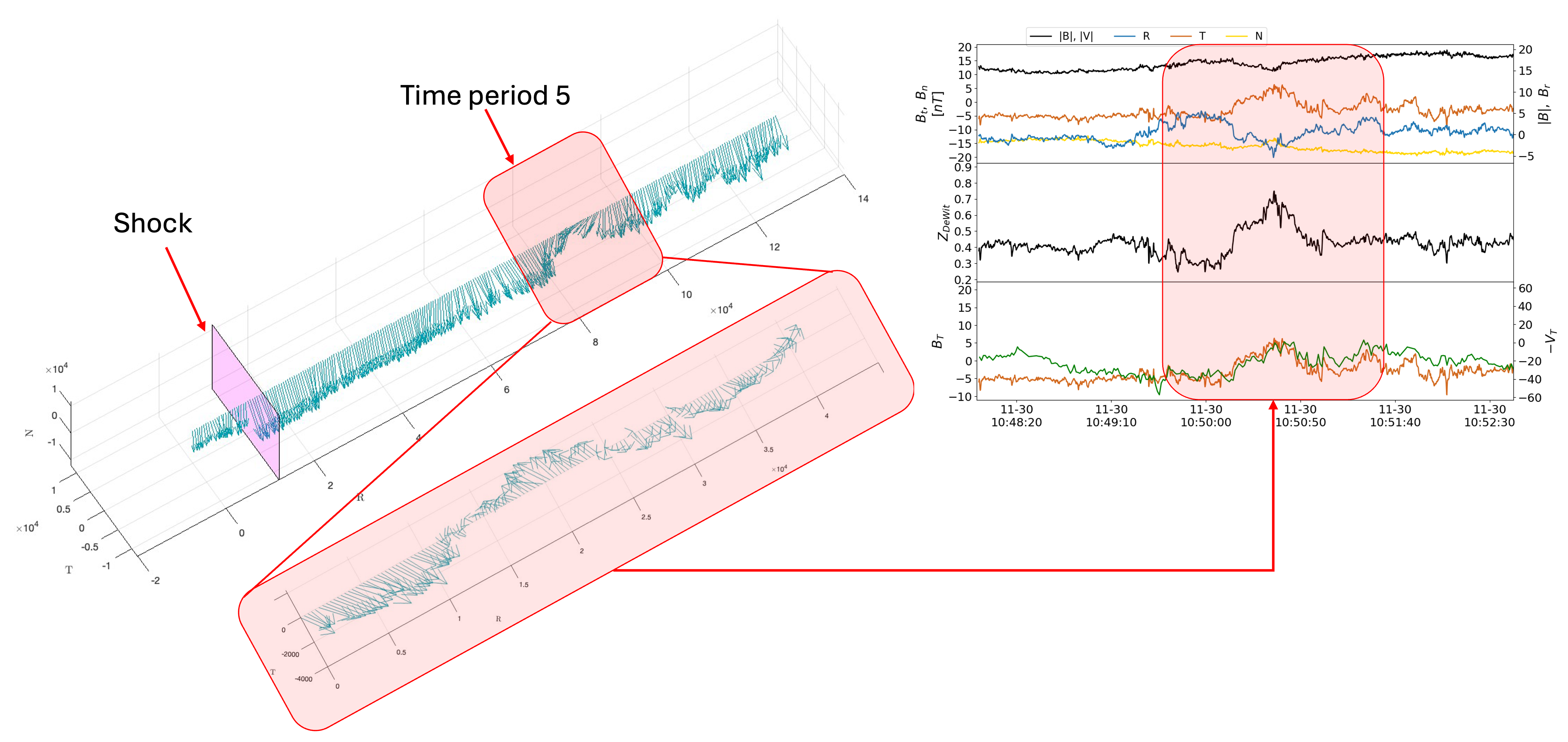}
    \caption{Left: 3D plot of the IMF vector in RTN coordinates in the immediate downstream of shock \#2 (magenta plane), where time period 5 is highlighted (red shaded area). Right: In-situ analysis targeting time period 5, where we display, from top to bottom, magnetic field magnitude and its components, the $Z$ parameter, and the $B-V$ correlation for the tangential components of magnetic field and ion velocity, indicating a fluctuation with a high degree of Alfv\`enicity and a cross helicity close to -1.}
    \label{fig:conveyor-belt}
\end{figure*}

Time period 5) warrants a more detailed discussion. 
The IMF surrounding this remarkable time period is shown in a 3D representation where magnetic field vectors have been plotted in space assuming that plasma moves at the average downstream speed (fig.~\ref{fig:conveyor-belt}, note that the units are arbitrary for illustration purposes). The figure shows the field change at shock \#2 (magenta plane), and it can also be seen that the IMF fluctuations are low in the immediate shock downstream until a the sudden change of the IMF conditions is seen corresponding to time period 5 (red shaded area, corresponding to  the excursions in the azimuth direction around 10:51:30 and 10:51:45 in fig.~\ref{fig:detail-shock-2}). Such rotation of the field, introduced in Section~\ref{subsec:TP5}, is clearly visible from the left-hand side of the Figure, where a zoom with a different viewpoint around it is also shown for clarity. We further characterized the magnetic field rotation by computing the Z parameter \citep{dudok-dewit-etal-2020}. As one can see from the right hand side of fig.~ \ref{fig:conveyor-belt} the rotation presents a mild decrease of the magnetic field magnitude, a high value of Z and a high degree of correlation between the magnetic and velocity fields.
These three features suggest that we are dealing with a magnetic switchback \citep[see, e.g.,][]{Kasper2019Natur.576..228K, dudok-dewit-etal-2020, larosa2021A&A}.

We note that, albeit modeling efforts have shown that magnetic switchbacks can affect energetic particle dynamics~\citep{Malara2023A&A}, whether magnetic switchbacks produce any observable features in energetic particle detection has been elusive so far. Our results, as presented in fig.~\ref{fig:detail-shock-2}) and fig.~\ref{fig:conveyor-belt} (time period 5) show that switchbacks can drive extremely dynamic, dispersive behavior in energetic protons at $\sim$ 100 keV energies. As other studies carried out with NASA's Parker Solar Probe did not find such an effect~\citep{Bandyopadhyay2021A&A}, we underline that such observations are only possible due to the unprecedented time and energy resolution achieved by Solar Orbiter EPD. 

The observations in time period 5) can be explained if the pitch-angle distribution is strongly centered around 90 degrees (i.e., pancake-like). The changing direction of the IMF with respect to the four EPT telescopes then results in such a pattern. As this structure (likely a switchback) is convected past Solar Orbiter, the high-energy particles bound to this structure with their larger gyro-radii take longer to be convected past the observer than lower-energy particles. This explains the apparent "inverse velocity dispersion".


\section{Conclusions}
\label{sec:concl}

It was the best of times, it was the worst of times, \ldots - for the thermal and suprathermal particle populations in the vicinity of this pair of very dissimilar shocks observed by Solar Orbiter at 0.83 au from the Sun. The first shock was a quasi-parallel shock  and showed strong non-equilibrium plasma signatures as seen in fig.~\ref{fig:om-2}. Because it ran into a small ICME with its low level of IMF fluctuations, suprathermal particles were not reflected back into the shock which therefore did not have an appreciable effect on energetic particles. The second, quasi-perpendicular shock exhibited unusual, rapid ion dynamics which were nevertheless well resolved with the EPD sensors aboard Solar Orbiter. Many of the variations of the energetic-particle fluxes were observed to be at the ion cyclotron time scale or even more rapid. The high time and energy resolution offered by Solar Orbiter's in situ instruments is critical for understanding the micro-physics of particle dynamics and acceleration at traveling interplanetary shocks. The rapid variations and dispersive features would not have been visible with the lower resolution provided by similar instruments on previous missions. They would probably have been interpreted as broad diffusive signatures.


\begin{acknowledgements}
     Solar Orbiter is a mission of international cooperation between ESA and NASA, operated by ESA. This work was supported by the German Federal Ministry for Economic Affairs and Energy and the German Space Agency (Deutsches Zentrum für Luft- und Raumfahrt, e.V., (DLR)), grant number 50OT2002. The UAH team acknowledges the financial support by Project PID2023-150952OB-I00 funded by MICIU/AEI/10.13039/501100011033 and by FEDER, UE. Solar Orbiter post-launch work is supported by NASA contract NNN06AA01C at JHU/APL and 80GFSC25CA035 at SwRI. Solar Wind Analyser (SWA) data are derived from scientific sensors which have been designed and created, and are operated under funding provided in numerous contracts from the UK Space Agency (UKSA), the UK Science and Technology Facilities Council (STFC), the Agenzia Spaziale Italiana (ASI), the Centre National d’Etudes Spatiales (CNES, France), the Centre National de la Recherche Scientifique (CNRS, France), the Czech contribution to the ESA PRODEX programme and NASA. Solar Orbiter magnetometer operations are funded by the UK Space Agency (grant ST/X002098/1). Work at IRAP is supported by CNRS, Centre National d’Etudes Spatiales (CNES) and the University of Toulouse (UT). R.K. acknowledges the postdoctoral fellowship from CNES. Solar Orbiter SWA work at UCL/MSSL is currently funded under UKSA/STFC grants ST/X002152/1 and ST/W001004/1. RCA acknowledges support from NASA grants 80NSSC21K0733, 80NSSC24K0908, and 80NSSC22K0993. L.W. acknowledges support from NSFC under contracts 42225404, 42127803 and 42150105. 
\end{acknowledgements}

\bibliographystyle{aa} 
\bibliography{tale} 

\appendix

\section{Determination of the critical Mach number}
\label{app:critical}

The concept of shock criticality is well--defined when shocks are considered thorough a fluid approach, where a critical Mach numbers arises when the downstream flow speed equals the downstream sound speed and a shock solution is only possible when some other dissipation process is introduced~\citep{Burgess2015}. In the collisionless case, such process is mediated by ion reflection, and the definition of the critical Mach number becomes somewhat vague, motivating several studies investigating supercriticality in different contexts~\citep[e.g.,][]{Amano2010,Ha2018}. The critical Mach number is determined by three quantities, the upstream $\langle \theta_{Bn} \rangle$, plasma $\beta$, and the ratio of specific heats, $\gamma$, of the solar wind plasma \cite[]{edmiston-and-kennel-1984}. The upstream values of the first two quantities for the two shocks considered in this paper are given in tab.~\ref{tab:tab_event}. The value of the third quantity, $\gamma$, cannot be measured directly and requires further consideration. The solar wind consists primarily of protons and doubly-ionized helium ions with a small admixture of heavy ions which, together with electrons, all contribute to the solar wind ratio of specific heats. \cite{de-avillez-etal-2018} investigated a plasma with solar composition and found that $\gamma$ is not constant, but depends on composition, temperature, and even the exact shape of the velocity distribution functions of the plasma constituents. The solar wind is depleted in helium \cite[]{geiss-1982, von-steiger-etal-2000} compared to the Sun, and is further depleted in the slow solar wind in which the two shocks were embedded \cite{bochsler-2007}. Using a lower limit of $v_{\rm th} = 10$ km/s for the thermal speed of protons (see the corresponding panels in figs.~\ref{fig:om-1} and \ref{fig:P2}) we arrive at $T>10^5$ K. Figure 4 of \cite{de-avillez-etal-2018} shows clearl that we can safely assume $\gamma \approx 5/3$ for these solar wind conditions. Using this value, we determined the critical Mach number, $M_C = 1.7$ ($M_C = 2.2$), for shock \#1 (\#2) by interpolating in fig.\,4 of \cite{edmiston-and-kennel-1984}. Thus we independently confirmed our conclusion that they were both supercritical based on the presence of the foot in both shocks \cite{bavassano-catteneo-etal-1986}.

\section{STEP measurements during time period 1)}
\label{app:step}

\begin{figure*}
    \centering
    \includegraphics[width=\textwidth]{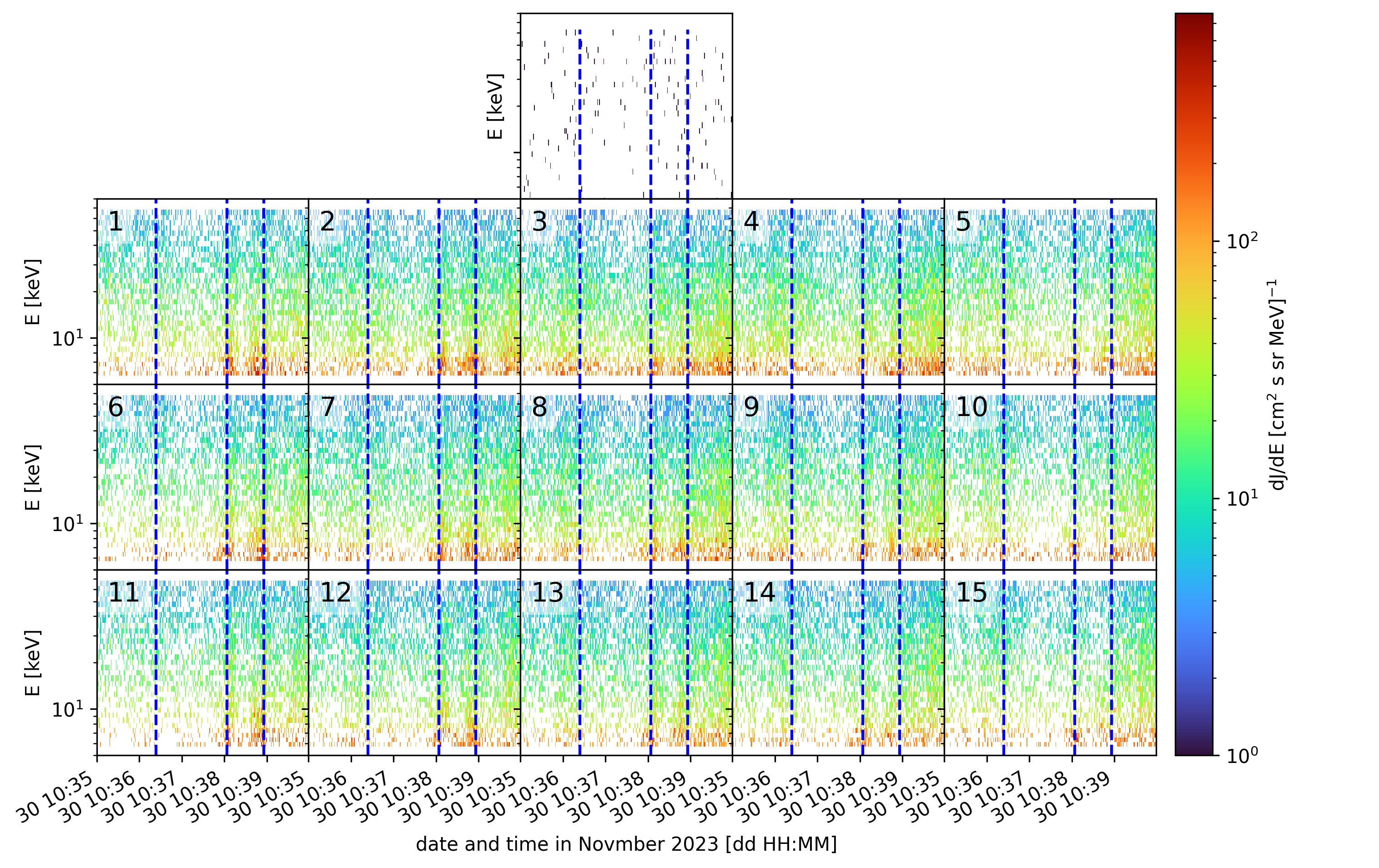}
    \caption{Detailed STEP measurements of time period 1) that complement fig.~\ref{fig:P2.1} show a very anisotropic distribution. The three vertical dashed blue lines in each panel denote the onset of the magnetic structure, the onset of the inverse velocity dispersion feature , and the end of the magnetic structure. The central pixel at the top shows background measurements, the numbers of the individual pixels are indicate in their upper left corners.}
    \label{fig:STEP_TP1}
\end{figure*}

\end{document}